\documentstyle[11pt,aaspp4,epsf]{article}

\begin{document}

\title{Oxygen abundances in unevolved metal-poor stars from near-UV  OH
lines\footnote{Based on observations  obtained with the William Herschel
Telescope, operated on the island  of La Palma by the Isaac Newton Group in the
Spanish Observatorio  del Roque de los Muchachos of the Instituto de Astrof\'\i
sica de  Canarias, as well as on observations made with the Anglo-Australian 
Telescope, Siding Spring, Australia.}}

\author{Garik Israelian, Ram\'on J. Garc\'\i a L\'opez, and Rafael Rebolo}
\affil{Instituto de Astrof\'\i sica de Canarias, \\ E-38200
La Laguna, Tenerife, \\ Spain} 

\authoraddr{Instituto de Astrof\'\i sica de Canarias, E-38200 La Laguna, 
Tenerife, SPAIN}

\authoremail{gil@ll.iac.es, rgl@ll.iac.es, and rrl@ll.iac.es}

\begin{abstract}
We have performed a detailed oxygen abundance analysis of 23 metal-poor 
($-$3.0$<$[Fe/H]$<-$0.3) unevolved halo stars and one giant through  the OH
bands in the near UV, using high-resolution echelle spectra.  Oxygen is found
to be overabundant with respect to iron in these stars, with the [O/Fe] ratio
increasing from 0.6 to 1 between [Fe/H]=$-$1.5 and $-$3.0. The behavior of the
oxygen overabundance  with respect to [Fe/H] is similar to that seen  in
previous works  based on \ion{O}{1} IR triplet data (Abia \& Rebolo 1989;
Tomkin et al. 1992; Cavallo, Pilachowski, \& Rebolo 1997). Contrary to the
previously accepted  picture, our oxygen abundances, derived from
low-excitation OH lines, agree well  with those derived from high-excitation
lines of the triplet. For nine stars in common with Tomkin et al. we obtain a
mean difference of 0.00$\pm$0.11 dex with respect to the abundances determined
from the triplet using the same stellar parameters and model photospheres.  For
four stars in our sample we have found measurements of the [\ion{O}{1}]
$\lambda$  6300 \AA\ line in the literature,  from which we derive oxygen
abundances consistent (average difference 0.09 dex) with those based on OH
lines, showing that the long standing controversy between oxygen abundances
from forbidden and permitted lines in metal-poor unevolved stars can be
resolved. Our new oxygen abundances show a smooth extension of the Edvardsson
et al.'s (1993) [O/Fe] versus metallicity curve to much lower abundances, with
a slope $-0.31\pm 0.11$ (taking into account the error bars in both oxygen
abundances and metallicities) in the range $-$3$<$[Fe/H]$<-$1. The
extrapolation of our results to very low metallicities indicates that the first
Type II SNe  in the early Galaxy provided oxygen to iron ratios 
[O/Fe]$\gtrsim$1. The oxygen abundances  of unevolved stars when compared with 
values  in the  literature for giants of similar metallicity imply that the
latter may  have suffered a process of oxygen depletion. As a result, 
unevolved metal-poor stars shall be considered better tracers of the early 
evolution of oxygen in the Galaxy.  The higher [O/Fe] ratios we find in dwarfs 
has an impact on the age determination of globular clusters, suggesting that
current age estimates have to be reduced by about 1--2 Gyr.

\end{abstract}


\section{Introduction}
 
Oxygen is the most abundant element in stars after H and He.  It is produced in
massive stars and ejected into the interstellar medium when they explode as
Type II SNe (see, for example,  \cite{arn78}; \cite{tin79}; \cite{wwo95}).
Iron, however, is produced in both Type II and less massive Type I SNe
(resulting from mass accretion by a C--O white dwarf, \cite{tin79}).  The fact
that Type I SNe have longer lifetimes offers an opportunity to put some
constraints on the time-scale of oxygen and iron  production by measuring
oxygen overabundance over iron.  It has been argued (\cite{tin79}) that at
least a 10$^{8}$-yr delay between Type I and Type II SNe can lead to an
enhancement of the [O/Fe] ratios in stars formed in very early epochs  of the
Galaxy. Theoretical models of Type II SN element yields show an increase in 
[O/Fe] (as well as other $\alpha$ elements) ratio as a function of the
progenitor mass (\cite{wwo95}).  Abundance ratios observed in low-mass
metal-poor stars allow estimates of primordial abundances of elements as well
as the  study of the yields of SNe of different progenitor masses.

Oxygen abundances  have been measured in many globular clusters and metal-poor
field stars  using: 1) the \ion{O}{1} IR triplet located at $\lambda$ 7774
\AA\  (\cite{abrl89}; \cite{tom92}; King 1994; \cite{cpr97});  2) the
[\ion{O}{1}] lines at $\lambda\lambda$ 6300, 6363 \AA\   (\cite{bar88};
\cite{gro89}; \cite{sne91}; \cite{kraf92}; \cite{kraf95}; \cite{min96}), and 3)
the OH lines in  the UV (Bessell, Sutherland, \& Ruan 1991; Nissen et al.
1994).  The use of different methods has a simple physical rationale. The
high-excitation ($\chi$ = 9.15 eV) triplet at $\lambda$ 7774 \AA\  is strong
and can be accurately measured in hot halo dwarfs, while forbidden lines are a
good tool for cool halo giants. Molecular  OH lines in the UV are observed in
both giants and dwarfs with $T_{\rm eff} \leq$ 6500 K (OH is destroyed at
higher temperatures). The most relevant  result from these studies was the
confirmation of the theoretically expected oxygen overabundance with respect to
iron. All authors agree on the point that [O/Fe] increases when [Fe/H]
decreases from 0 to $-$1. However, the slope of the [O/Fe] versus [Fe/H]
dependence and its behavior for metallicities below $-$1 were not clear.
Results of different authors disagree. Studies of [\ion{O}{1}] data suggest an
[O/Fe] overabundance of +0.4 to +0.5 dex for [Fe/H] $\leq -$0.8 (\cite{bar88};
\cite{gro89}; \cite{sne91}; \cite{kraf92}), while the \ion{O}{1} triplet gave
systematically higher values;  [O/Fe] $\sim$ +0.79$\pm$0.29 (see Cavallo et al.
1997, and references therein).

Abia \& Rebolo (1989) derived  oxygen abundances from the \ion{O}{1}  IR
triplet of 30 dwarfs and subgiants covering a range of metallicities $-$2.7
$\leq$ [Fe/H] $\leq$ 0 and found that [O/Fe] reaches  1.0--1.2 at about
[Fe/H]$\sim -2$. It became clear that the [\ion{O}{1}] line at 6300 \AA\ give
systematically lower abundances compared with the \ion{O}{1} IR triplet.
Different ideas appeared in the literature aiming to  resolve this discrepancy.
King (1993) has suggested a new temperature scale for metal-poor dwarfs
increasing the effective temperatures by 150--200 K. However, the validity of
the assumptions made in his temperature scale were criticized  by Balachandran
\& Carney (1996). A possible source of  systematic errors could be hidden in
the form of the source function of the oxygen triplet. It was clear that
forbidden lines and IR triplet are relatively  free from non-LTE effects in
metal-poor stars (\cite{kis93}; Tomkin et al. 1992). Kiselman \& Nordlund
(1995) proposed that correct radiative transfer in 3-D atmospheric models can
decrease the strength of the \ion{O}{1} IR triplet relative to the
[\ion{O}{1}]. Takeda (1995) has addressed this problem in terms of the
excitation temperature effects after finding that the solar  \ion{O}{1} IR
triplet weakens when the chromosphere is  taken into account in the
computations. However, it is not obvious how to apply the ideas of Takeda
(1995) and  Kiselman \& Nordlund (1995) to the metal-poor stars.  Recent
observations and analysis by  Cavallo et al. (1997) have shown that the
problem  remains unsolved. 

Most of the measurements of the forbidden lines of oxygen have been performed
in giants. Barbuy (1988) and Graton \& Ortolani (1989) have reported
[O/Fe]=0.35$\pm$0.15 and [O/Fe]=0.5$\pm$0.1 from 20 and 18 field halo giants,
respectively.  Sneden et al. (1991) and Kraft et al. (1992) derived
[O/Fe]=0.36$\pm$0.05 and [O/Fe]=0.34$\pm$0.02 for 10 and 16 globular cluster
giants with [Fe/H]$< -1.9$ and $-1.3 >$[Fe/H]$> -1.9$,  respectively. These and
other studies show  O overabundances of [O/Fe]$\simeq$0.4 down to metallicities
of [Fe/H]$\simeq -3$. In a given cluster,  giants with  almost the same
metallicity  may show differences in the oxygen abundance of a factor 2--3 (and
even 10) with respect to other  (Kraft et al. 1995; Kraft 1994,  and references
therein). As regards metal-poor dwarfs, the analysis of [\ion{O}{1}]  lines has
been limited to those of moderately low metallicity because of the low strength
of these transitions. Only a few measurements have been reported by Spite \&
Spite (1991), Spiesman \& Wallerstein (1991) and Nissen \& Edvardsson (1992)
showing consistent results with previous findings in giants.

An alternative way of deriving oxygen abundances from  OH  lines in the UV was
suggested by Bessell, Hughes, \& Cottrell (1984). These lines are very strong
and can be measured at much lower abundance levels than is possible with the
forbidden or the permitted \ion{O}{1} lines. However, this measurement 
requires the use of near-UV high-resolution spectroscopy, which for cool and
distant very metal-poor stars can only be obtained in large-diameter
telescopes. In addition, there are a number of difficulties  related to the
specific problems of deriving abundances from the lines located in the near
UV.  Even with the help of synthetic spectra, the identification of many blends
is uncertain. In many cases it is hard to separate a weak line from the noise
or a blend. In order to overcome this difficulty one needs to obtain high-S/N
and high-dispersion spectra. This method has been used by Bessell et al. (1991)
and Nissen et al. (1994) who derived oxygen abundances in eight and nine 
metal-poor stars, respectively. Bessell et al. (1991)  noticed that  the
low-excitation lines of OH ($\chi$ = 0--1.7 eV) will provide us with more
accurate abundances (compared  with the IR triplet) since these lines are
formed in those layers of the atmosphere where the majority of neutral metal
lines used for the abundance studies occur. The oxygen abundances derived from
the OH lines were about 0.5 dex lower than those measured from the  \ion{O}{1}
triplet. Thus, the abundances derived from OH and [\ion{O}{1}] agreed, and it
was natural to conclude that something was wrong with the  IR triplet.  

In the present paper, we analyze new OH data and derive oxygen abundances in a
larger sample conformed by 23 metal-poor unevolved halo stars and one giant. We
compare our abundances with those from OH, the \ion{O}{1} IR triplet, and
[\ion{O}{1}] for stars in common with previous works and discuss the
implications of our findings.

\section{Observations}

These observations belong to a wider project aimed at measuring beryllium,
oxygen, nitrogen, and carbon abundances in metal-poor stars by using atomic and
molecular lines located in the near UV. The observations were carried out in
different runs using the Utrecht Echelle Spectrograph (UES) at the Nasmyth
focus of the 4.2-m William Herschel Telescope of the Observatorio del Roque de
los Muchachos (La Palma), and the UCL Echelle Spectrograph (UCLES) of the 3.9-m
Anglo-Australian Telescope. Most of the spectra were obtained using the UES
with a resolution of $R=\lambda /\Delta\lambda\sim 50000$, while the resolution
achieved with the UCLES was $R\sim 60000$. A detailed log of the observations,
including telescopes, dates, and exposure times can be found in Garc\'\i a
L\'opez et al. (1998).

All the spectra were reduced using standard {\sc iraf} \footnote{{\sc iraf} is
distributed by the National Optical Astronomical Observatories, which is
operated by the Association of Universities for Research in Astronomy, Inc.,
under contract with the National Science Foundation, USA.} procedures (bias
subtraction, flatfield correction, and extraction of one-dimensional spectra).
Different spectra for each object were co-added before wavelength calibration
and continuum normalization. The  final signal-to-noise (S/N) ratio varies for
the different echelle orders,  being in the range 30--50 for most of the stars.
The  final spectra span  from 3025 to 3430 \AA\ for the UES data, and from 
3070 to 3390 \AA\ for the UCLES, including some gaps between the orders.

\section{Abundance analysis}

\subsection{Stellar parameters}

The 24 stars analyzed in this paper are listed in Table 1. Effective
temperatures ($T_{\rm eff}$) were estimated initially using the Alonso,
Arribas, \& Mart\'\i nez-Roger (1996) calibrations versus $V$$-$$K$ and
$b$$-$$y$ colors, which were derived applying the infrared flux method
(Blackwell et al. 1990), and cover a wide range of spectral types and metal
content. All the details on photometric measurements are provided by Garc\'\i a
L\'opez et al. (1998). The selection of metallicities for the program stars was
carried out by critically inspecting a large number of published values, and
the  references can also be found in that work.   

Surface gravities ($\log g$) were determined by comparing the observed
Str\"omgren $b-y$ and $c_1$ indices with synthetic ones generated using the
corresponding filter transmissions and a grid of Kurucz (1992) blanketed model
atmospheres fluxes. This is also explained in detail by Garc\'\i a L\'opez et
al. (1998), where a comparison is given with the spectroscopic estimates by
other authors. In particular, Nissen, H$\o$g, \& Schuster (1998) have recently
derived surface gravities for a sample of 54 metal-poor stars using {\it
Hipparcos} parallaxes to determine luminosities. For ten stars in common, the
gravities derived by Garc\'\i a L\'opez et al. are slightly lower
systematically, with a mean difference of $0.22\pm 0.13$ dex in $\log g$. As we
shall see below, a difference of this magnitude affects the oxygen abundance
derived from OH lines by about 0.07 dex.

Garc\'\i a L\'opez et al. (1998) computed synthetic spectra in the region
around the \ion{Be}{2} doublet at $\lambda$ 3131 \AA\ (used for deriving
beryllium abundances) for different combinations of stellar parameters, within
the error bars provided by the photometric calibrations; Table 1 lists the
values which best reproduce the observations. These values also reproduce well
other bluer and redder spectral regions where the OH lines used in this paper
are located. In particular, the effective temperatures assigned to our sample
agree well with those used by Abia \& Rebolo (1989) and Tomkin et al. (1992),
with mean differences of $-3\pm 41$ K (7 objects in common) and $+13\pm 53$ K 
(9 objects in common), respectively. The mean difference between our  $T_{\rm
eff}$ values and those used by King (1993; who proposed a hotter temperature
scale) is $-146\pm 57$ K for 9 stars in common. For other works with a smaller
number of common objects with our sample, like those of Spite \& Spite (1991)
and Nissen et al. (1994), the differences in $T_{\rm eff}$ are within our
adopted error bars. These comparisons and the fact that a change of 100 K in
$T_{\rm eff}$ translates into a change of $\sim 0.2$ dex in the derived oxygen
abundance (as we will show below), limit strongly the abundance differences
with previous works which could be attributed to the adopted temperature
scale.

\subsection{Atomic and molecular data}

Atomic data were obtained from the VALD database (Piskunov et al. 1995). Line
lists of CH, NH and other molecules were obtained from the CD-ROM no. 15 of
Kurucz (1993). Laboratory wavelengths for the OH lines were taken from Stark,
Brault, \& Abrams (1994), and line parameters (oscillator strengths, $\log gf$,
and lower excitation potential, $\chi$) from Goldman \& Gillis (1981). The OH
lines of the solar spectrum were inspected  using the high-resolution solar
flux atlas of Kurucz et al. (1984) and the Moore, Minnaert, \& Houtgast (1966)
atlas to identify relatively strong unblended OH lines. However, independently
of how careful we were in selecting the OH lines, this strategy does not
guarantee complete success due to the large number of unidentified lines in the
solar spectrum.  We have avoided an important number of strong OH lines because
they were strongly blended with other metals for which we do not have accurate
$gf$ values. There were several unblended candidate OH lines in the regions
$\lambda\lambda$ 3105$\pm$10 and 3150$\pm$5 \AA, but these were unfortunately 
located between our echelle orders. With due regard for possible blends, we
have selected 8 clean and 3 weakly blended unsaturated features corresponding
to the same spectral range available for our stellar observations. These are
mainly OH (0,0) lines of the A$^2\Sigma$-X$^2\Pi$ electronic system in the
spectral region 3080--3300 \AA, and are listed in Table 2. The only OH line in
common with Nissen et al. (1994) is $\lambda$ 3139.165 \AA. This is a very
strong line blended with a weak \ion{Fe}{1} line at $\lambda$ 3138.903 \AA\
($\log gf$=$-$2.5, $\chi$=3.05 eV) which disappears at metallicities below
[Fe/H]$\sim -0.5$.  We have also found that all the other lines used by Nissen
et al. (1994) were either blended or located between our echelle orders (note
that these authors used only lines located in the spectral region 3138--3155
\AA). Most  of the  OH lines in our list are strong enough to be found in
metal-poor stars with [Fe/H] $\leq$ $-$2. Two OH lines at $\lambda\lambda$
3128.060 and 3128.107 \AA\ are blended, giving rise to a single OH feature, and
the OH line at $\lambda$ 3128.286 \AA\ is blended with the weak \ion{Sc}{2}
$\lambda$  3128.269 \AA\ ($\log gf$=$-$0.08, $\chi$=3.46 eV).

To minimize the effects associated with the errors in the transition
probabilities of atomic and molecular lines in this region, we adjusted the
$\log gf$ values of the lines used until we succeeded in reproducing the solar
atlas of Kurucz et al. (1984) with solar abundances. This procedure is, in
principle, strictly valid only for stars with characteristics similar to those
of the Sun, but it has been shown (e.g. Garc\'\i a L\'opez, Severino, \& Gomez
1995) that it provides consistent results in abundance analyses carried out for
metal-poor stars in the near UV. For the OH lines in particular, this approach
is justified so long as  solar $gf$ values are insensitive to the treatment of
damping. This is indeed the case for most of the stars in our sample, for which
we have been able to fit the OH lines without assuming any ad hoc values for
Van der Waals broadening (two exceptions associated with stars of high gravity
and low $T_{\rm eff}$ will be discussed separately).

\subsection{Spectral synthesis}

We have employed a grid of LTE, plane-parallel, constant flux, and blanketed
model atmospheres provided by Kurucz (1992, private communication), computed 
with {\sc atlas9} without overshooting, which are interpolated for given values
of $T_{\rm eff}$, $\log g$, and [Fe/H]. Synthetic spectra were obtained using
the LTE code {\sc wita3}, a new UNIX-based version of code {\sc abel6}
(Pavlenko 1991). 

Synthetic solar spectra were computed in the neighborhood of the selected OH
lines, and were convolved with a Gaussian of FWHM =0.052 \AA\ to
reproduce the instrumental profile of the solar atlas. We used a model with
$T_{\rm eff}$=5777 K, $\log g$=4.4, [Fe/H]=0.0, microturbulence $\xi=1~{\rm
km}~{\rm s}^{-1}$, and solar chemical abundances taken from Anders \& Grevesse
(1989). The solar oxygen abundance used was log $N$(O)=8.93 (on
the customary scale in which log $N$(H)=12). Values of $gf$ for the OH and
other lines  where adjusted to reproduce the solar spectrum. Theoretical and
adjusted (i.e. solar) $gf$-values of the OH lines are listed in Table 2, and
examples of the fits obtained in two spectral regions are shown in Figure 1. We
did not find a systematic difference between the theoretical and adjusted $\log
gf$ values for the (0,0) lines used by us, as was the case for Nissen et al.
(1994), who lowered the oscillator strengths of these lines by 0.16 dex to fit
the solar spectrum.

Using this set of solar $gf$ values we proceeded to derive the oxygen
abundances of our stars by computing synthetic spectra with the stellar
parameters listed in Table 1, and changed the oxygen abundance until a good fit
to the observations was obtained. Synthetic spectra were convolved with a
Gaussian with an adequate FWHM to reproduce the resolutions achieved in the UES
and UCLES spectra, respectively. Table 3 lists the abundances derived from each
line considered with respect to the solar abundance ([O/H]=log
(O/H)$_\star$--log (O/H)$_\odot$), and the mean value from all lines. The lack
of data in the table corresponds mainly to the bluest region where the
signal-to-noise ratios were the lowest for most of the stars, as well as to
those cases in which a combination of low S/N and weak lines did not allow us
to derive the abundance. HD 166913 is the worst case with only two lines. It
also appears that the unblended $\lambda$ 3255 \AA\ line located in a high-S/N
region gives the lowest abundances for several stars, but we have no reasons to
discard it a priori. Figure 2 shows different examples of comparisons between
synthetic and observed spectra.

Two stars of the sample, HD 64090 and HD 103095, with low effective temperature
and high surface gravity, show observed OH line profiles which are narrower
than the corresponding synthetic ones. The instrumental profile associated with
their spectra is the same  as that corresponding to other stars observed with
the UES spectrograph, and for which we have obtained a good fit to the observed
lines. To improve the fit, Ya. V. Pavlenko kindly included an option in {\sc
wita3} to compute damping constants for a few molecules (including OH), which
was applied to the analysis of these two stars. The resulting synthetic spectra
were much closer to the observed ones with this improvement, although in
general they were still a little poorer than those obtained for the rest.   

\subsubsection{Uncertainties}

To check the sensitivity of the derived oxygen abundances to the input stellar
parameters, we computed the changes in several of the strongest OH lines
associated with variations of the stellar parameters  for a set of model
atmospheres ($T_{\rm eff}$=5000, 5600, and 6200 K, $\log g$=3 and 4, and
[Fe/H]=$-0.5$ and $-2$). It turned out that a typical change of $+100$ K in
$T_{\rm eff}$ increases [O/H]  by $\sim +0.2$ dex. Surface gravity works in the
opposite direction, and a change of $+0.3$ dex in $\log g$ corresponds to a
change of $\sim -0.1$ dex in [O/H], whereas an error in metallicity of 0.3 dex
implies an uncertainty of $\sim 0.05$ dex in [O/H]. The same results have been
obtained by Nissen et al. (1994) using {\sc osmarcs} model atmospheres 
(Edvardsson et al. 1993), and a different line list. In our syntheses we used a
fixed value for the microturbulence,  $\xi=1~{\rm km}~{\rm s}^{-1}$, and we
confirm the previous finding by Nissen et al. (1994), which indicated that
errors in the microturbulence and in the differential abundance, [C/Fe], can be
neglected. The uncertainties associated with the stellar parameters were
estimated for each star by taking these values into account and combining them
in quadrature. 

Systematic errors, like unrecognized line blends and the continuum placement,
also arise in this process. The location of the continuum constitutes a  
serious problem for low-S/N UV data. These errors vary from star to star. In
many cases it is impossible to distinguish between weak spectral features and
noise. The situation is especially complex for stars with metallicities
[Fe/H]$\sim -1.2\pm0.3$. Spectra of stars with [Fe/H]$\ge -1$ can be compared
with the Sun in order to make an accurate differential analysis, while the
continuum can be placed more easily in the spectra of very metal-poor stars due
to the disappearance of a large number of weak metallic lines. A measure of
these uncertainties can be obtained from the dispersion (standard deviation) in
abundances found from different OH lines, which are listed in Table 2. In a
typical case (with S/N$\sim 30-40$ and based on 9 lines) the error is about
0.09 dex. The final errors listed in Table 1 take into account the
uncertainties from the stellar parameters and the dispersion in abundances from
different OH lines.

Another systematic error comes from the comparison between our adopted
gravities and those based on {\it Hipparcos} measurements. As mentioned above,
the $\log g$ values adopted in this work appear to be systematically lower by
$\sim 0.2$ dex. To put our oxygen abundances on the {\it Hipparcos} scale we
should decrease the [O/H] values listed in Table 2 by $\sim 0.07$ dex (given
the sensitivity of our analysis to changes in surface gravity).

A final check on the uncertainties related to the use of a specific set of
model atmospheres and spectral synthesis code was performed by computing
synthetic spectra using {\sc osmarcs} models and the LTE code {\sc spectrum}
(included in the Uppsala Synthetic Spectrum Package). {\sc osmarcs} models were
kindly provided by B. Edvardsson. Below $\lambda$ 4500 \AA\ they are computed
using opacity sampling techniques, while at longer wavelengths opacity
distribution functions are employed. The latter method is used for computing
Kurucz models at all wavelengths. The differences in oxygen abundances from
{\sc osmarcs} models with 5795/4.30/$-0.45$ ($T_{\rm eff}$/$\log g$/[Fe/H]),
6210/4.00/$-2.25$, and  5550/3.35/$-2.70$ with respect to their corresponding
Kurucz values are in the range $-0.09$ to $-0.15$ dex.

\section{Comparison with previous oxygen measurements}

\subsection{Comparison of oxygen abundances derived from OH lines}

The oxygen abundance of HD 140283 has been previously derived from OH lines by
Bessel \& Norris (1987), Ryan, Norris, \& Bessell (1991), Bessell et al.
(1991), and Nissen et al. (1994). This star  deserves special attention for its
low metallicity and high-S/N UV  spectrum (it is a bright metal-poor star with
$V=7.2$ mag). Figure 3 shows the  comparison between the synthetic and observed
profiles for this star corresponding to various regions surrounding our
selected OH lines. Synthetic spectra were computed using the stellar parameters
listed in Table 1 (5550/3.35/$-$2.7) and show that [O/H] is lying between
$-1.9$ and $-1.8$ (the mean value resulting from all the  OH lines in Table 3
is [O/H]$=-1.85$), while a value as low as $-2.1$ (obtained by Nissen et al.
1994) does not fit the observations. Abundances derived from the OH lines at
$\lambda$ 3085 \AA\ are more uncertain due to the low S/N achieved in the
bluest echelle order. However, the local continuum can be accurately placed in
the neighborhood of the lines located between 3100--3300 \AA\ because of the
high S/N of the spectra and the low metallicity of this star. Assuming
5540/3.5/$-$2.7 from Nissen et al. (1994), we arrive at [O/H]$=-1.9$ from the
strongest OH lines in our list ($\lambda$ 3139.169 \AA), just 0.1 dex lower
than the corresponding abundance listed in Table 3, which is not surprising due
to the high similarities between the set of stellar parameters used in both
cases.

Furthermore, one of the sets of parameters used in the comparison between
synthetic spectra computed with Kurucz and {\sc osmarcs} models corresponds to
the values adopted for HD 140283. In this case the oxygen abundance derived
using the {\sc osmarcs} model is 0.15 dex smaller than that obtained from the
Kurucz one. The oxygen abundance corresponding to this star would be then
[O/H]$=-2.0$. This means that the oxygen abundance derived by us, using our
line list and an {\sc osmarcs} model atmosphere, is indeed only about 0.1 dex
higher than that obtained by Nissen et al. (1994), and this difference is
directly associated with the small differences in $T_{\rm eff}$ and $\log g$
between both analyses.  

Assuming 5779/3.79/$-$2.49 from Nissen, H{\o}g, \& Schuster (1998), where
$T_{\rm eff}$ was estimated using the Alonso, Arribas, \& Mart\'\i nez-Roger
(1996) calibration based on $b$$-$$y$, and $\log g$ is derived from the
parallax measured by {\it Hipparcos}, we arrive at [O/H]$=-1.6$. This is
because $T_{\rm eff}$ has a larger effect on the strength of the OH lines than
$\log g$. However, given the metallicity  adopted by Nissen et al., the [O/Fe]
ratios obtained with these parameters and with those listed in Table 1 will
differ by 0.06 dex only. 

Bessell et al. (1991) used the following set of parameters for HD 140283:
5700/3.3/$-$2.8 and  5800/4.5/$-$2.8. The high $\log g$ value was derived from
a previous parallax which is 2.5 times larger than the value measured by {\it
Hipparcos}, and we have not taken it into account. Using 5700/3.3/$-$2.8 we
obtain [O/H]$=-1.5$, which is 0.35 dex larger than the abundance derived from
our adopted parameters and would imply also a larger [O/Fe] ratio. The
difference of 1 km s$^{-1}$ in $\xi$ between both analyses has a negligible
influence on the abundances derived. Bessell et al. (1991) used {\sc marcs}
model atmospheres from a grid provided by  Bell, Gustafsson, \& Eriksson (1979,
private communication), which have much less UV opacity and different overall
spectral shapes as compared to the {\sc osmarcs} and Kurucz models (see Fig. 2
of Edvardsson et al. 1993), so the agreement between their oxygen abundance
([O/H]$=-1.8$), and our value at $-1.85$ for this star is just a coincidence
due to the stellar parameters and abundance analysis working in opposite
directions.  

Assuming 5750/3.2/$-$2.7 and 5600/3.0/$-$2.8, Ryan et al. (1991) arrived at
[O/H]$=-2.1$ and $-2.3$, which corresponds to [O/Fe]$=+0.6$ and $+0.5$,
respectively. These values translate into [O/H]$=-1.4, -1.7$ and [O/Fe]$=+1.3,
+1.1$, respectively, with our analysis. The model atmospheres used by these
authors were those of Bell (1981, private communication) and Kurucz (1979),
which also differ strongly in the UV from the {\sc osmarcs} and Kurucz (1992)
models. Finally, Bessell \& Norris (1987) derived [O/H]$=-1.75$ and
[O/Fe]$=+0.9$  for a model with 5650/3.5/$-$2.65 (the corresponding values
obtained by us are [O/H]$=-1.8$ and [O/Fe]$=0.9$). The line list used in that
work was based on the Goldman \& Gillis (1981) data. This is also the case in
our analysis, as well as in those of  Bessell et al. (1991) and Nissen et al.
(1994), but not that of Ryan et al. (1991), who used a line list based on
Kurucz (1989, private communication).

Another star in common with Nissen et al. (1994) is HD 84937. Using the set of
parameters 6090/4.0/$-$2.4 they obtained [O/H]=$-$1.87. Adopting our stellar
parameters (Table 1) and taking into account the difference between Kurucz and
{\sc osmarcs} models their abundance translates into [O/H]=$-$1.57, which is 
just slightly lower than our mean value listed in Table 3 for five OH lines
($-$1.45$\pm$0.07).

In summary, independent LTE synthetic spectra of near-UV OH lines obtained
using {\sc osmarcs} or Kurucz (computed with the code {\sc atlas9} without
overshooting) model atmospheres, line lists taken from Goldman \& Gillis (1981)
and adjusted to reproduce the solar spectrum, and the same stellar parameters
(especially $T_{\rm eff}$) provide very similar oxygen abundances. These
abundances are shown to be more reliable than those obtained when using older
model photospheres with a coarser treatment of the opacities in the UV.

\subsection{Oxygen abundances from OH versus IR triplet}
 
To compare our OH oxygen abundances with those coming from the \ion{O}{1} IR
triplet, we have computed them using the equivalent widths measured by Tomkin
et al. (1992) and the stellar parameters listed in Table 1. The same LTE code,
model atmospheres, and microturbulence used for the OH lines were employed,
while the oscillator strengths were taken from Tomkin et al. Our solar oxygen
abundance, derived using 5777/4.44/0.0 and equivalent widths of the \ion{O}{1}
triplet measured on the Kurucz et al. (1984) solar atlas, is log $N$(O)$=8.98
\pm 0.03$ (the error quoted is associated only with the uncertainties in
measuring the equivalent widths). Using a microturbulence $\xi=1.5$ km s$^{-1}$
(instead of 1 km s$^{-1}$) would decrease the solar abundance to log
$N$(O)$=8.92$. In Fig. 7 are plotted the oxygen abundances with respect to the
Sun derived from OH lines and the IR triplet for nine stars in common with our
sample. The agreement between both sets of abundances is very good (with a mean
difference of $0.00\pm 0.11$ dex) and they delineate the same trend of [O/H]
with [Fe/H]. An independent comparison with abundances derived from the
\ion{O}{1} triplet is shown in Fig. 8, where we have plotted the abundances
derived from OH lines together with those derived (in LTE) directly by Tomkin
et al. (1992), and Cavallo et al. (1997). It can be seen that, in general, the
[O/H] and [O/Fe] ratios follow the same trend with metallicity for both 
indicators. The abundances derived from the OH lines are mainly located in the
lower part of the distribution of the measurements in the diagrams, which
can be explained in terms of the different stellar parameters and models
of atmospheres used by the other authors as we have seen above. 

As an external confirmation to these comparisons, Spite (1997) has shown that
it is possible to reconcile the oxygen abundances derived from the \ion{O}{1}
triplet and UV OH lines in dwarf stars by carefully selecting the effective
temperatures and using state-of-the-art model photospheres. She illustrates
this with a re-analysis of three stars ($-2>$[Fe/H]$>-3$) studied by Tomkin et
al. (1992), and Nissen et al. (1994), for which Spite et al. (1996) provide
$T_{\rm eff}$ values based on a $T_{\rm eff}$$-$$(b$$-$$y)_0$ relation. Once
the reddening has been taken into account, this relation provides effective
temperatures which are systematically 40--60 K hotter than those adopted by
Tomkin et al. Spite (1997) estimated the new values of [Fe/H], [O/Fe]$_{\rm
OH}$, and [O/Fe]$_{\rm O I}$, and find $-2.2$, 0.66, and 0.68, respectively,
which correspond to the mean values of the three stars. This [O/Fe]$_{\rm OH}$
value is $\sim 0.2$ dex lower than our mean value in the same metallicity
range, which, again, can be explained  by the effects of using Kurucz and {\sc
osmarcs}  model photospheres  (the latter one employed by Spite) and slightly
different stellar parameters.

Balachandran \& Carney (1996) studied the oxygen abundance of the star HD
103095 in detail using IR OH and CO lines located at the $H$ and $K$ bands,
respectively; they also consistently re-assessed the abundances derived from 
the \ion{O}{1} triplet and the [\ion{O}{1}] $\lambda$ 6300 \AA\ line with
similar model atmospheres to those employed in our work. With a set of stellar
parameters also very similar to those used in our analysis, they obtained
[O/H]$_{\rm OH,CO}=-0.93$, [O/H]$_{\rm O I}=-0.80$, and [O/H]$_{\rm [O
I]}=-0.89$. This star is included in the group of nine stars in common with
Tomkin et al. (1992), and our estimate for the oxygen abundance derived from
the \ion{O}{1} triplet, [O/H]$_{\rm O I}=-0.84$, is in excellent agreement with
the corresponding value of Balachandran \& Carney. However, there is a
difference of 0.24 dex between the abundances derived from OH lines in the UV
and IR. As explained before, HD 103095 is also one of the two stars for which
the synthetic OH spectra look  broader than the observed ones, and this could
explain part of the discrepancy. In any case, the difference in abundance is
well within the error bars. 

Tomkin et al. (1992) examined both C and O in halo dwarfs, using
$_{\rm C I}$ and CH lines and the O triplet.
They found a temperature-dependent divergence between the abundances 
from \ion{C}{1} and CH and concluded that the \ion{C}{1} abundances suffered 
a temperature-related error. They have also argued by extension that the 
\ion{O}{1} abundances suffered a similar error, but that the stellar C/O ratio 
could be reliably obtained from the \ion{C}{1} and \ion{O}{1} lines. Based on 
this, they concluded that the C/O ratio in halo dwarfs was relatively constant. 
There is a problem when making the analogy between \ion{C}{1} versus CH and 
\ion{O}{1} versus OH. In the present paper we do not find a divergence between 
abundances for \ion{O}{1} and OH. We consider that the fact that oxygen and 
carbon abundances derived from molecular lines do not show trends versus 
effective temperature, in contrast with the abundances derived from atomic 
lines, could be an indication of model atmosphere and radiative transfer
problems which deserve a careful investigation which goes beyond the
scope of this paper."

\subsection{OH versus forbidden lines}

We have found only four dwarfs in our sample for which oxygen abundances have
been derived using [\ion{O}{1}]. These are HD 22879 (King \& Boesgaard 1995),
HD 76932 (Magain 1987; Barbuy 1988), HD 103095 (Spite \& Spite 1991;
Balachandran \& Carney 1996), and HD 134169 (Spite \& Spite 1991). In Table 4
we list the model parameters used by each of these authors and the oxygen
abundances they inferred from the [\ion{O}{1}] line (columns 3 and 4,
respectively). In order to compare these abundances with those we have derived
from OH lines we have synthesized the forbidden oxygen line adopting the same
set of stellar parameters than for the OH analysis and the $gf$ value given
by Lambert (1978). We derived the solar oxygen abundance from the $\lambda$
6300 \AA\ forbidden line measured on the Kurucz et al. (1984) solar atlas and
using the same model atmosphere previously employed for the OH and \ion{O}{1}
lines. From an equivalent width of 5.4 m\AA\ we obtained a solar oxygen
abundance log $N$(O)$=8.90$, which is consistent with similar analyses in the
literature (see e.g. Balachandran \&  Carney  1996). Then, using the equivalent
widths provided by the authors for the 6300 \AA\ line and our model parameters
(Table 1) we derived the abundances that are listed in column 5, which we find
in reasonable agreement with those in column 4, but still lower than those we
derived from our OH lines (column 6). 

When we compared the sensitivity of OH lines and the [\ion{O}{1}] $\lambda$
6300 \AA\ line to the stellar parameters we noticed the high effect of gravity
in the latter case. We decided to investigate whether the reason of the
discrepancy may lie in an inappropriate gravity determination for these
objects. Fortunately, {\it Hipparcos} parallaxes are available for these stars
from which we have inferred new, presumably better, gravities listed in column
7. In order to derive these gravities we followed the same recipe than Nissen,
H{\o}g, \& Schuster (1998). These values are larger by 0.22 dex in average than
the values we adopted in Table 1, which implies a reduction of the oxygen
abundance inferred from OH lines and an increase of the abundance derived from
the forbidden line. The final values are listed in columns 8 and 9 of Table 4,
respectively, where both abundances show very good agreement for  HD 22879 and
HD 134169 and still some reasonable discrepancy of $\sim$ 0.15 dex for the
other two stars, which can be accounted for by the uncertainties associated
with the equivalent width measurements, which are particularly difficult in the
case of extremely weak lines as it is the case of the forbidden line in these
two stars (equivalent width less than 2 m\AA). In addition, and as we have
already mentioned before, our OH analysis for HD 103095 provides broader than
observed lines which make difficult to carry out detailed comparisons for this
star. The average difference between oxygen abundances derived from OH and
[\ion{O}{1}] lines for our four stars is 0.09 dex. This exercise strongly
suggests that an {\it Hipparcos}-based gravity scale may indeed be key to
explain the discrepancies on oxygen abundances from forbidden and permitted
lines in several unevolved metal-poor stars.

\section{Discussion}

\subsection{Dependence of [O/H] and [O/Fe] on metallicity}

The [O/H] abundances listed in Table 1 show a clear decrease with decreasing
metallicity, and this is illustrated in Figure 4. A noteworthy feature of this 
figure is the smooth continuation of the trend found by Edvardsson et al.
(1993) from their study of disk stars with $-$1.0 $<$[Fe/H]$<$0.3 using the
\ion{O}{1} IR triplet. Note that their abundances were scaled to the
[\ion{O}{1}] results of Nissen \& Edvardsson (1992). Abundances derived 
from OH lines are slightly higher than those of Edvardsson et al. Simple 
linear regressions (not taking the error
bars into account) performed on both samples show a common slope of 
$0.63\pm0.03$, and a systematic difference of $0.05\pm 0.06$ dex in 
the abundances. This effect could be associated with the systematic 
differences observed in abundances derived from Kurucz and {\sc osmarcs} 
model photospheres.
 
Our [O/Fe] ratios are listed in Table 1 and plotted versus [Fe/H] in Figure 5. 
A shift of $+0.05$ dex has been applied to the values of Edvardsson et al. 
(1993) to move them, on average, to the same abundance scale as that of our
analysis. The figure shows a smooth trend of increasing [O/Fe] with decreasing
metallicity. Nissen et al. (1994) found an increase in [O/Fe] from 0.4 to 0.8
between [Fe/H]=$-$1.5 and $-$3.0. The corresponding increase in Fig. 5  goes
from 0.6 to 1.0. As explained in \S 3, a difference of $\sim 0.2$ dex is
related to the model photospheres and stellar parameters used in both analyses.
The data do not indicate any clear flattening out of [O/Fe] with decreasing
metallicity below [Fe/H]$\sim -1$, and can be fitted using a straight line with
a slope $-0.33\pm 0.04$ (not taking the error bars into account) in this
metallicity range. Taking into account the error bars in oxygen abundances and
metallicities, using the routine {\it fitexy} of Press et al. (1992) available
in the IDL Astronomy User's Library, the resulting slope is $-0.31\pm 0.11$ for
17 stars.  

We have investigated the presence of possible systematic dependences of [O/H] 
on $T_{\rm eff}$ or $\log g$ in Fig. 6 (upper and lower panels, respectively).
The absence of any trend in these figures suggests that the uncertainties in
the derived oxygen abundances are only related to the errors listed in Table 1
(with the exception of the errors implicit in the model photospheres and the
uncertainties in applying the solar $gf$ values to stars with different
characteristics from the Sun).

\subsection{Comparison between dwarf and giant stars: mixing in giants}

We have discussed above the reliability of oxygen abundances derived  from  OH
lines  in metal-poor dwarfs and we are driven to conclude that the [O/Fe] ratio
is systematically higher in dwarfs than in giants (up to 0.5 dex). Does this
mean that O is depleted in giants, or is this discrepancy related to problems
in the analysis of giants or in the analysis of the forbidden lines in general?

It is well known that the analysis of chemical abundances in cool metal-poor
giants is subject to significant uncertainties. For instance, we note the work
by Dalle Ore (1993), who found serious inconsistencies among the temperatures
obtained from the excitation and ionization equilibria of several cool giants.
The reliability of abundances based on the [\ion{O}{1}] $\lambda$ 6300 \AA\ has
been questioned by King \& Boesgaard (1995). Therefore, there are reasons to be
cautious when inferring conclusions from these measurements in giants and it
would be worth measuring different oxygen spectral lines in these stars.
Nevertheless, it is possible that the oxygen content of metal-poor giants is
really lower on average than in dwarfs of similar metallicity. A
plausible reason for this is mixing of nuclear-processed material, depleted in
oxygen, with the upper atmospheric layers.  Evidence for oxygen depletion in
globular-cluster giants, associated with N enhancements, has been found by 
Pilachowski (1988), and   low ratios of $^{12}$C/$^{13}$C and $^{12}$C/$^{14}$N
have been  discovered in several metal-poor giants (Sneden, Pilachowski, \& 
VandenBerg 1986), indicating depletion of carbon. A global  anticorrelation
between O and Na has been discovered by Cohen (1978) and Peterson (1980) and
confirmed by more recent studies of giants in M13, M3, and M5 (Kraft et al.
1992, 1995; Kraft 1994; Minniti et al. 1996). Recently Pilachowski, Sneden,
\& Kraft (1996) compared Na  abundances in 60 metal-poor halo subgiants, giants
and  horizontal-branch stars with similar metallicities (but of higher 
luminosity) in globular clusters. They found a small effect of increasing
[Na/Fe] by 0.12 dex when moving from halo subgiants to horizontal-branch stars,
which could be  interpreted as due to the dredgeup of Na-enriched material from
CNO burning. In addition to CNO elements, Na, Mg and Al abundances are also
correlated in a large number of globular cluster stars (Kraft 1994) showing
the general effects of mixing. However, the degree of mixing which can be
appealed is restricted by observables like the $^{12}$C/$^{13}$C ratios, since
it is likely that ON-cycled material will also show very low ratios.

On the theoretical side, Sweigart \& Mengel (1979) have already proposed
meridional currents to cause the mixing of CNO products with the upper
atmosphere. It also became clear that Na and Al enhancements observed  in many
globular-cluster giants (Kraft 1994) can be produced by the
$^{22}$Ne(p,$\gamma$)$^{23}$Na reaction of the NeNa-cycle (Langer, Hoffman, \&
Zaidins 1997, and references therein). However, the fact that $^{22}$Ne is not
an abundant part of Ne suggests considering the 
$^{14}$N($\alpha$,$\beta$)$^{18}$F($\beta^{+}$,$\nu$)$^{18}$O($\alpha$,$\gamma$)$^{22}$Ne 
reaction as a source of $^{22}$Ne. Thus, one would need a material which had
undergone significant O$\rightarrow$N or C$\rightarrow$N processing. In the
O-depleted layers one would also expect a transformation of the $^{26}$Mg and
$^{25}$Mg isotopes of Mg into Al in the MgAl-cycle (Langer, Hoffman, \& Sneden
1993). Recently Denissenkov, Weiss, \& Wagenhuber (1997) argued against the
purely primordial interpretation (Cottrell \& Da Costa 1981) of the large
star-to-star  abundance variations observed in many globular-cluster giants.
They have proposed a mechanism which combines primordial abundance anomalies
with deep mixing. It has also been shown recently (Langer et al. 1997) that the
use of  updated rates of Ne--Na cycle reactions can explain the Na--O abundance
anticorrelation observed in globular-cluster giants, and the Na--N correlation
observed in field halo giants.  Even without going into details of the theory
which explains depletion of different elements and their mixing with an upper
atmosphere, it has become increasingly apparent that giants are not reliable 
for studying the primordial abundances of CNO-cycle elements. Clearly, it is
not safe to use CNO, Na, Al and Mg elements in giants as probes of Galactic
chemical evolution.

\subsection{On the chemical evolution of oxygen in the Galactic halo}

Our [O/Fe] versus [Fe/H] curve shows that the [O/Fe] ratio  increases
monotonically with decreasing metallicity until it reaches $\sim$ 1.0 at
[Fe/H]$\sim$$-3.0$. We do {\sl not} observe a  clear ``plateau'' or a change in
the slope at low metallicities when all error bars are taken into account, in
contrast with McWilliam et al.'s (1995) claim for a flattening in the
overabundance of other $\alpha$ elements at metallicities between $-$2.0 and
$-$3.0. The most metal-deficient star in our sample (BD $+$23\arcdeg 3130) is
a  giant, and it is possible that the [O/Fe] derived for this star is not
primordial for the reasons mentioned above, but, even removing this object from
the plot, the evidence for any flattening of the distribution of points at very
low metallicites is very marginal, and it seems that the [O/Fe] ratio continues
increasing below [Fe/H]$\sim$$-3$. It is of considerable importance  to extend
the number of oxygen measurements at such low metallicities to investigate the
nucleosynthesis by supernovae at early times, the mixing of the ejected 
material in the interstellar medium (Audouze \& Silk 1995), as well as the
possible existence of a pre-Galactic population of super-massive stars.

There is increasing evidence for a large scatter in the abundances of  other
$\alpha$ elements in some halo stars with metallicities below $-3$. McWilliam
et al. (1995) found negative [Mg/Fe] ratios in the stars CS22968-014 and
CS22952-015, with [Fe/H] near $-3.4$, while Fuhrmann, Axer, \& Gehren (1995)
found [Mg/Fe]$=-0.28$ for BD $+$3\arcdeg 740. At the same time, McWilliam et
al. (1995) have found [Mg/Fe]$=1.2$ for CS22949-037, another very metal-poor
star with [Fe/H]$=-3.99$. The origin of this scatter, and its possible relation
to the MgAl nuclear cycle (Langer et al. 1993),  which would induce large
enhancements of Al which are not observed, are issues which  still remain
unresolved. Star-to-star differences exist also for  heavy elements in
metal-poor stars (Gilroy et al. 1988;  McWilliam et al. 1995). This is in fact
an indication that the protohalo was not well mixed before these low-mass stars
formed. The large abundance variations observed at very low metallicities can
be caused, for example, by   shot noise from the small number of SNe
determining the Galactic nucleosynthesis during the first epochs  of star
formation.  It will be very important to check whether or not the oxygen
abundances show a similar scatter at very low metallicities, as well as the
correlation with Mg and Ca. This abundance determination can be carried out by
using OH lines, since they will be observable even at extremely low 
metallicities. 

\subsection{The ages of globular clusters}

Freedman et al. (1994) and  Mould et al. (1995) estimate the age of  the
Universe from measurements of the Hubble constant based on observations  with
{\it Hubble Space Telescope} ({\it HST}) and arrive at $t_{0}$=8--11 Gyr, while
Sandage et al. (1996) find $t_{0}$=11--15 Gyr. These  determinations appear to
be in conflict with the so far generally accepted  ages of globular clusters
$\tau$=15.8$\pm$2.1 Gyr (see e.g. Bolte \& Hogan 1995).   Recently D'Antona,
Caloi, \& Mazzitelli (1997) have shown that the FST  (full spectrum of
turbulence) models of stellar evolution may decrease the estimated ages of
globular clusters but by an amount not exceeding  10$\%$ with respect more
conventional stellar evolution models. Their best estimate is $\tau$=11--12
Gyr, offering a plausible way to alleviate any age discrepancy.

Our finding of the high [O/Fe] ratios at low metallicities may also  play an
important role in resolving an ``age conflict''. The abundance of oxygen
affects both the internal opacity and the energy generation  (thermonuclear
burning) of cool metal-poor  stars near the turn-off loci of the
color-magnitude diagram. Ages derived from comparison of theoretical
evolutionary tracks with  observational color-magnitude diagrams will be
significantly affected (VandenBerg 1992, and references therein). It has been
shown that, for very metal-poor stars, the reduction in  age caused by an
oxygen enhancement is first of all caused  by the change in the H-burning
nucleosynthesis rate through the CNO cycle. From the computations of isochrones
and the plots of the turn-off luminosity versus age (VandenBerg 1992) one finds
that an increase of 0.5 dex in [O/Fe] implies a  reduction of 1 Gyr in the
estimated age. If our oxygen values are representative of the initial  oxygen
abundances of globular-cluster stars, it would imply  a considerable reduction
of globular cluster ages.  For example, the use of isochrones with oxygen
enhancement given by our value [O/Fe]=0.8 at [Fe/H]=$-$2.5 will reduce the age
estimation of a globular cluster by 2 Gyr with respect to isochrones with solar
[O/Fe]. D'Antona et al. (1997) considered an oxygen enhancement of  0.3--0.5
dex when dealing with the age  analysis for the most metal-poor globular
clusters; based in  our results, a larger enhancement, 0.8--1.0 dex, appears
more adequate, implying  that their best age estimate for these clusters should
be reduced by 1 Gyr, to arrive at 10--11 Gyr. It is very important to 
determine oxygen abundances in turn-off stars of the most metal-poor globular
clusters to confirm whether their oxygen overabundances are as high as we find
in field unevolved halo stars.

\section{Conclusions}

We have obtained oxygen abundances of metal-poor unevolved stars from an
analysis of OH lines observed with high spectral resolution in the near
ultraviolet. We find that the [O/Fe] ratio increases linearly as metallicity
decreases from [Fe/H]$\sim $$-$0.5 to $-3$, reaching values of $\sim$ 0.6 at
[Fe/H] =$-$1.5 and $\sim$ 1.0 at [Fe/H] =$-$3.0. This result is in good
agreement with the [O/Fe] ratios derived using the \ion{O}{1} IR triplet (Abia
\& Rebolo 1989; Tomkin et al. 1992; Cavallo et al. 1997), and is consistent
with a smooth extrapolation of the results by Edvardsson et al. (1993) in disk
stars of higher metallicities. 

For two stars in common with the OH analysis of Nissen et al. (1994) we find 
oxygen abundances that differ by less than 0.2 dex. The difference, which is
within error bars, can be explained by the use of different model atmospheres
(Kurucz models versus {\sc osmarcs}), different line lists, and slightly
different  stellar  parameters. For four stars in our sample we have derived
oxygen abundances from the [\ion{O}{1}] $\lambda$  6300 \AA\ line which are
consistent within measurement errors with those inferred from the OH lines.
The use of presumably more reliable gravities determined from {\it Hipparcos}
parallaxes seems to be key  to understand the discrepancies  between abundances
based on these two types of lines in dwarf stars. The fact that oxygen
abundances derived from UV lines, from the \ion{O}{1} triplet, and from the
[\ion{O}{1}] line agree supports the reliability of our measurements. Our work
stresses the previously noticed  discrepancy between abundances of metal-poor
dwarf and giant stars. Either the analysis of the forbidden lines in giants is
subject to larger uncertainties than currently accepted, or oxygen is depleted
in these stars via mixing processes, which is rather plausible given the amount
of evidence  supporting internal mixing in globular-cluster giants and to some
extent in their field counterparts.
   
We conclude that field halo dwarfs are better tracers of the early evolution of
oxygen in the Galaxy. We do not find any flattening out of [O/Fe] with
decreasing  metallicity from 0 to $-$3.  Extrapolation of our results to much
lower metallicities suggests that the average ratio of oxygen to iron provided
by the first Type II SNe was close to 1.0 dex or possibly higher, as opposed to
previous findings based on the analysis of the forbidden lines in giants giving
values close to 0.5 dex.  Our observations suggest a reduction of current
estimates of ages of very metal-poor globular clusters which may contribute
significantly to remove any conflict among them and values of the age of the
Universe inferred from recent measurements of the Hubble constant.

\acknowledgments
We are grateful to Ya. V. Pavlenko for providing the code {\sc wita3} and for
helpful discussions. P. Nissen has kindly provided us with their unpublished
paper about {\it Hipparcos} data, C. Allende Prieto helped us in deriving the 
surface gravities for four stars from {\it Hipparcos} parallaxes, B. Gustafsson
sent us the OH line list used by Nissen et al. (1994), and B. Edvardsson kindly
computed several {\sc osmarcs} models. The comments from an anonymous referee
have been of value in improving the content of this article, especially in the
comparison between the abundances derived from OH and [\ion{O}{1}] lines.

This research has made use of the VALD database, and was partially supported by
the Spanish DGES under projects PB92-0434-C02-01 and PB95-1132-C02-01.

\clearpage

\begin{figure}[ht]
\epsscale{0.95}
\plotone{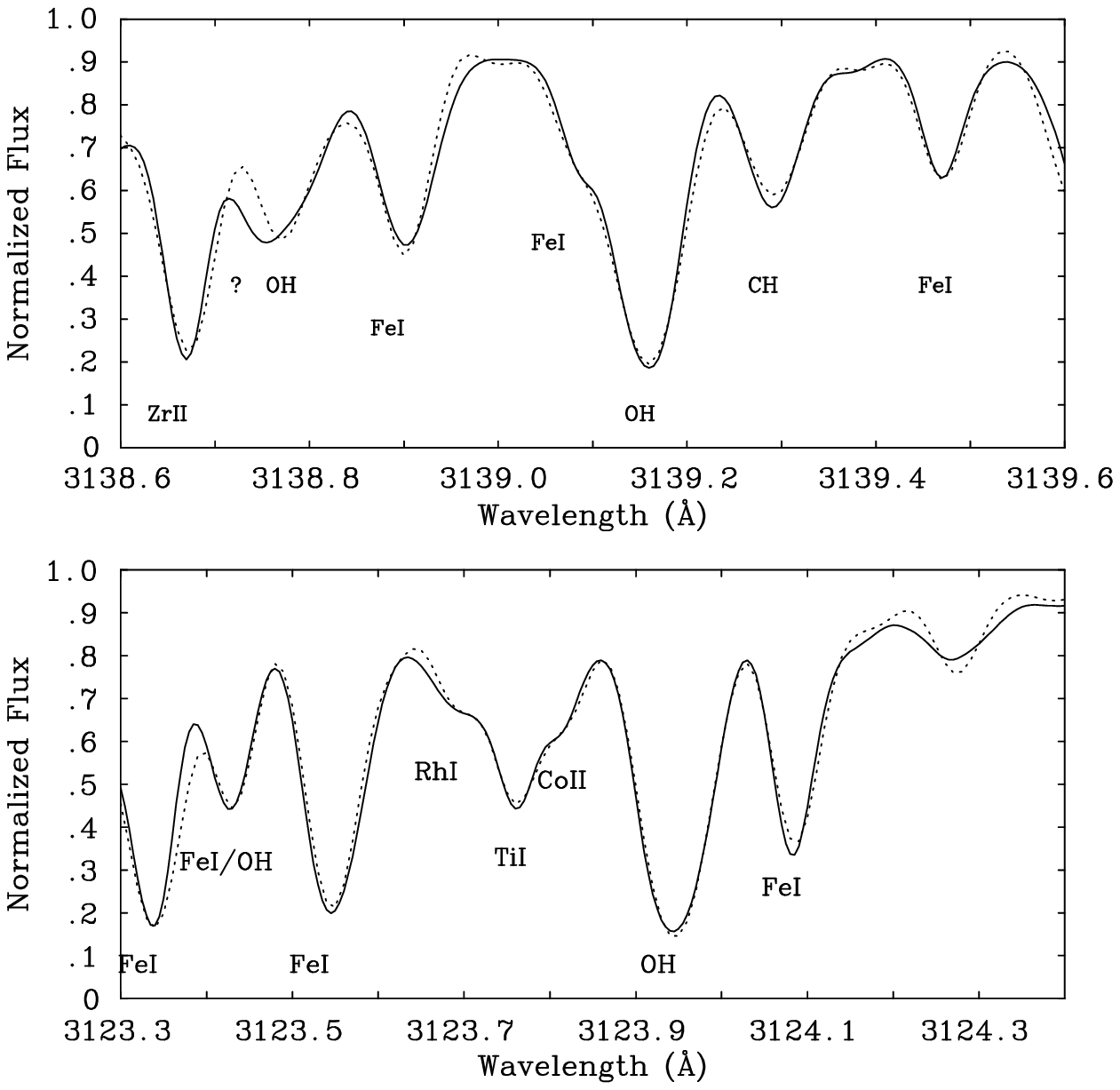}
\caption{Fit of synthetic spectra (dotted lines) to two regions
surrounding the OH lines in the high-resolution integrated-flux solar atlas of
Kurucz et al. (1984; solid lines). The locations of the different OH lines are
indicated while question marks are associated with cases without a clear line
identification. \label{ff1}}
\end{figure}

\begin{figure}[ht]
\epsscale{0.95}
\plotone{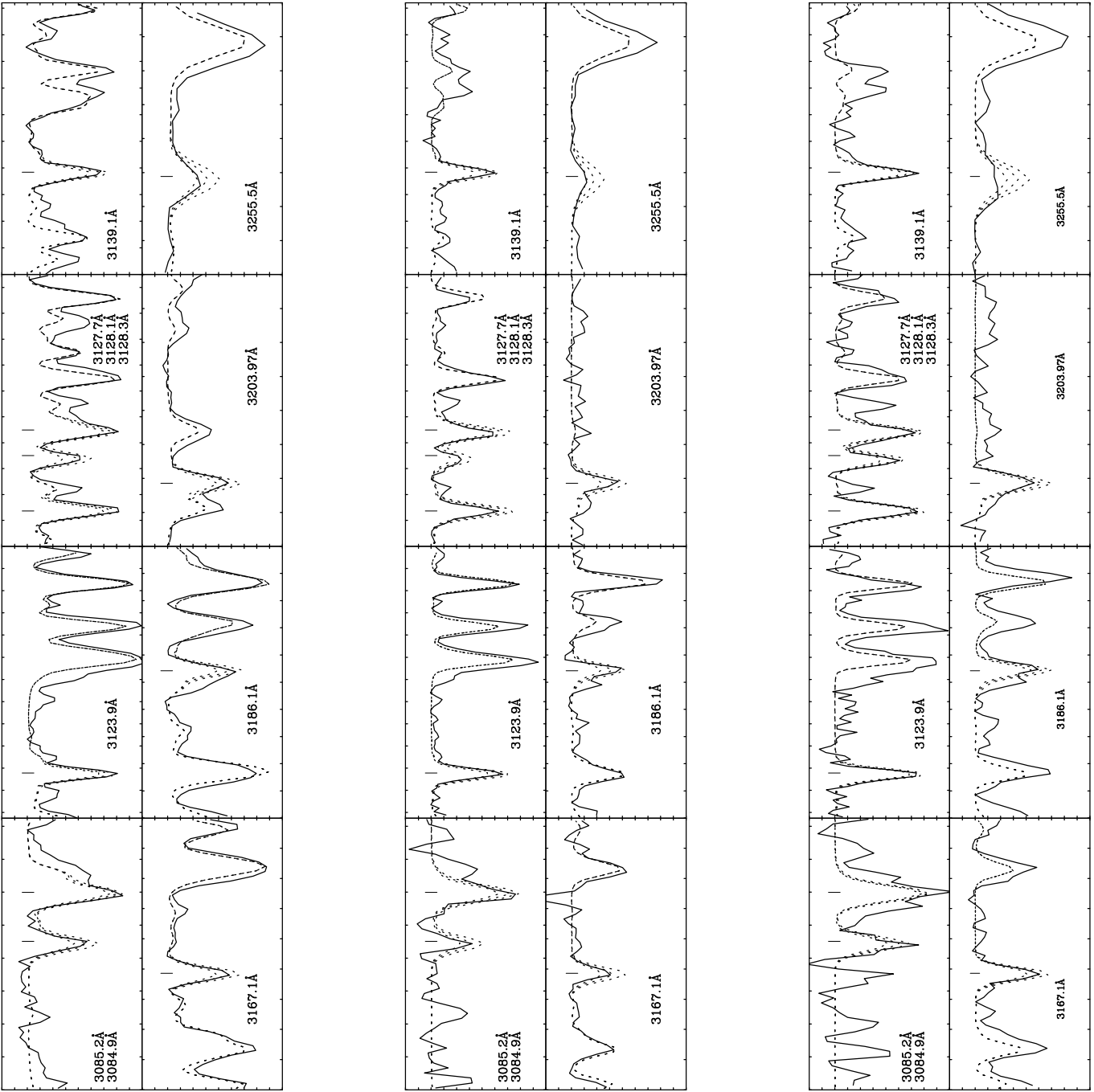}
\caption{Comparison between observed (solid lines) and synthetic 
spectra (dotted lines) for three  metal-poor stars: HD 76932 (upper panel), HD
19445 (middle panel) and BD $+$23\arcdeg 3130 (bottom panel). Synthetic spectra
were computed for three different oxygen abundances for each star (HD 76932,
[O/H]$=-0.9, -0.7, -0.5$; HD 19445, [O/H]$=-1.55, -1.35, -1.15$; BD
$+$23\arcdeg 3130, [O/H]$=-1.9, -1.7, -1.5$). All fluxes are normalized and the
$y$-axis has the same range from 0.1 to 1.2 in all plots.  \label{ff2}}
\end{figure}

\begin{figure}[ht]
\epsscale{0.95}
\plotone{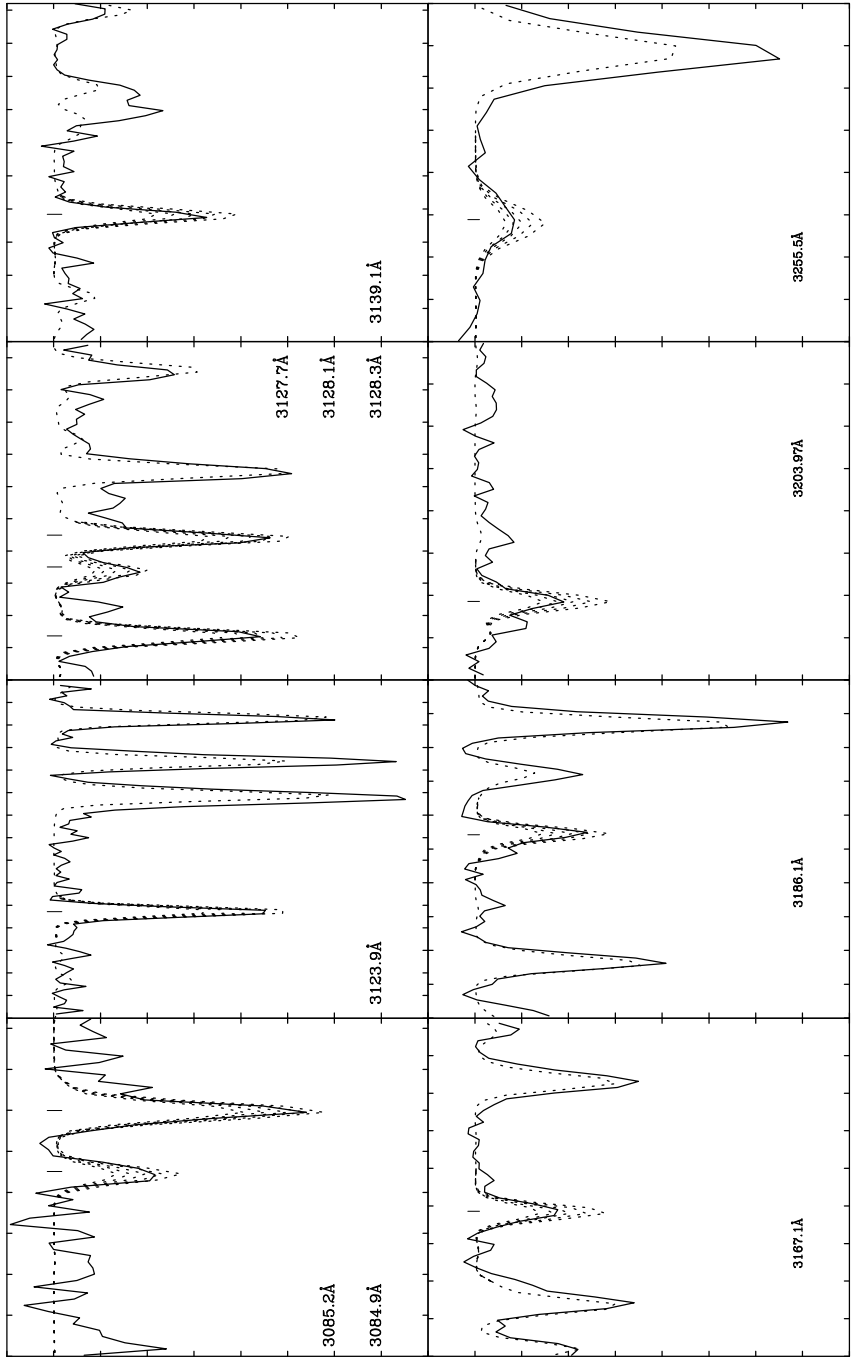}
\caption{Comparison between observed (solid lines) and synthetic 
spectra (dotted lines, corresponding to [O/H]=$-2.1, -2.0, -1.9, -1.8,$ and
$-1.7$) of HD 140283. All fluxes are normalized and the $y$-axis has the range
from 0.2 to 1.1 in all plots. \label{ff3}}
\end{figure}

\begin{figure}[ht]
\epsscale{0.8}
\plotone{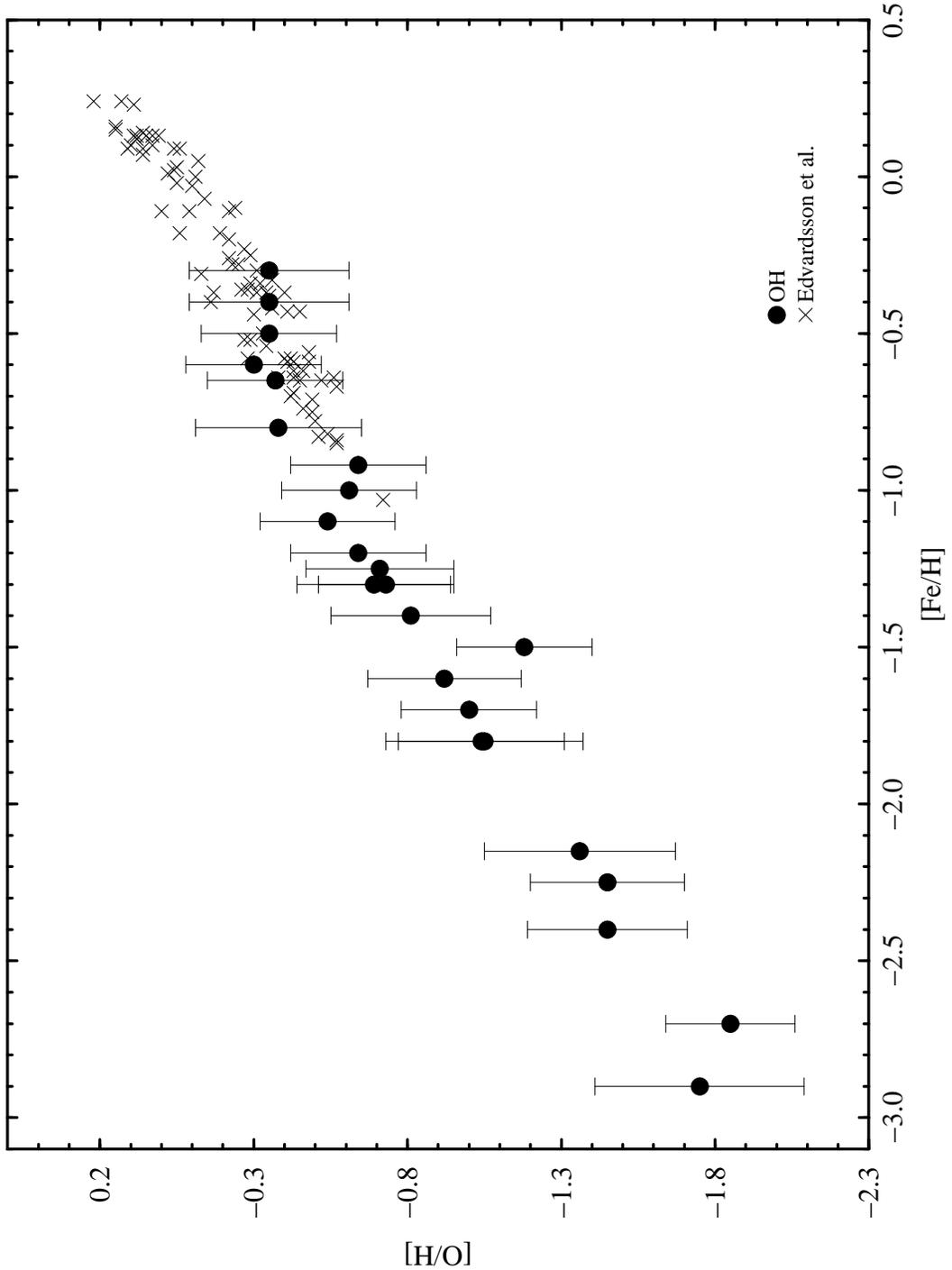}
\caption[]{Oxygen abundances (with respect to the Sun), derived from the
UV OH lines, against metallicity for the 24 stars analyzed in this article. 
Also included in the plot are the oxygen abundances derived by Edvardsson et
al. (1993) from the IR \ion{O}{1} triplet in metal-rich stars. The abundances
from OH lines show a smooth progression from metal-rich to metal-poor stars. A
mean difference of $\sim 0.05$ dex observed between the [O/H] scales of both
samples is explained by using Kurucz ({\sc atlas9} without overshooting) and
{\sc osmarcs} model atmospheres in the analyses of OH lines and the \ion{O}{1}
triplet, respectively.\label{Fig4n}}
\end{figure}

\begin{figure}[ht]
\plotone{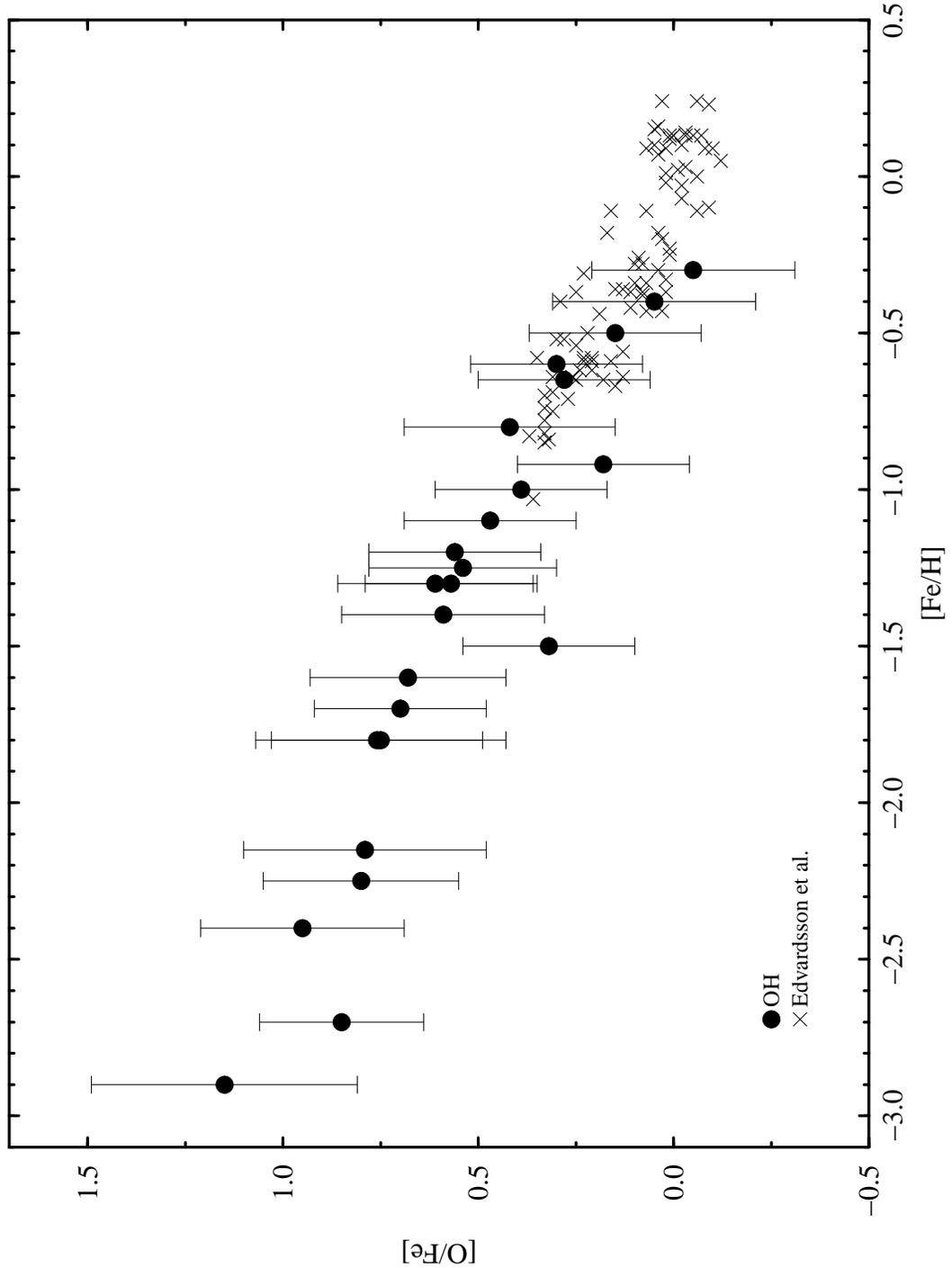}
\caption[Fig5.ps]{Oxygen overabundances (derived from OH lines) with respect to
iron against metallicity for the 24 stars analyzed in this article. Also
included in the plot are the values derived by Edvardsson et al. (1993) from
the IR \ion{O}{1} triplet but shifted by $+0.05$ dex to move them to the same
abundance scale than our analysis. Note the smooth continuous increase in
[O/Fe] with decreasing metallicity, which reaches a value $\sim +1$ for
[Fe/H]$=-3$.
\label{fig5}}
\end{figure}

\begin{figure}[ht]
\plotone{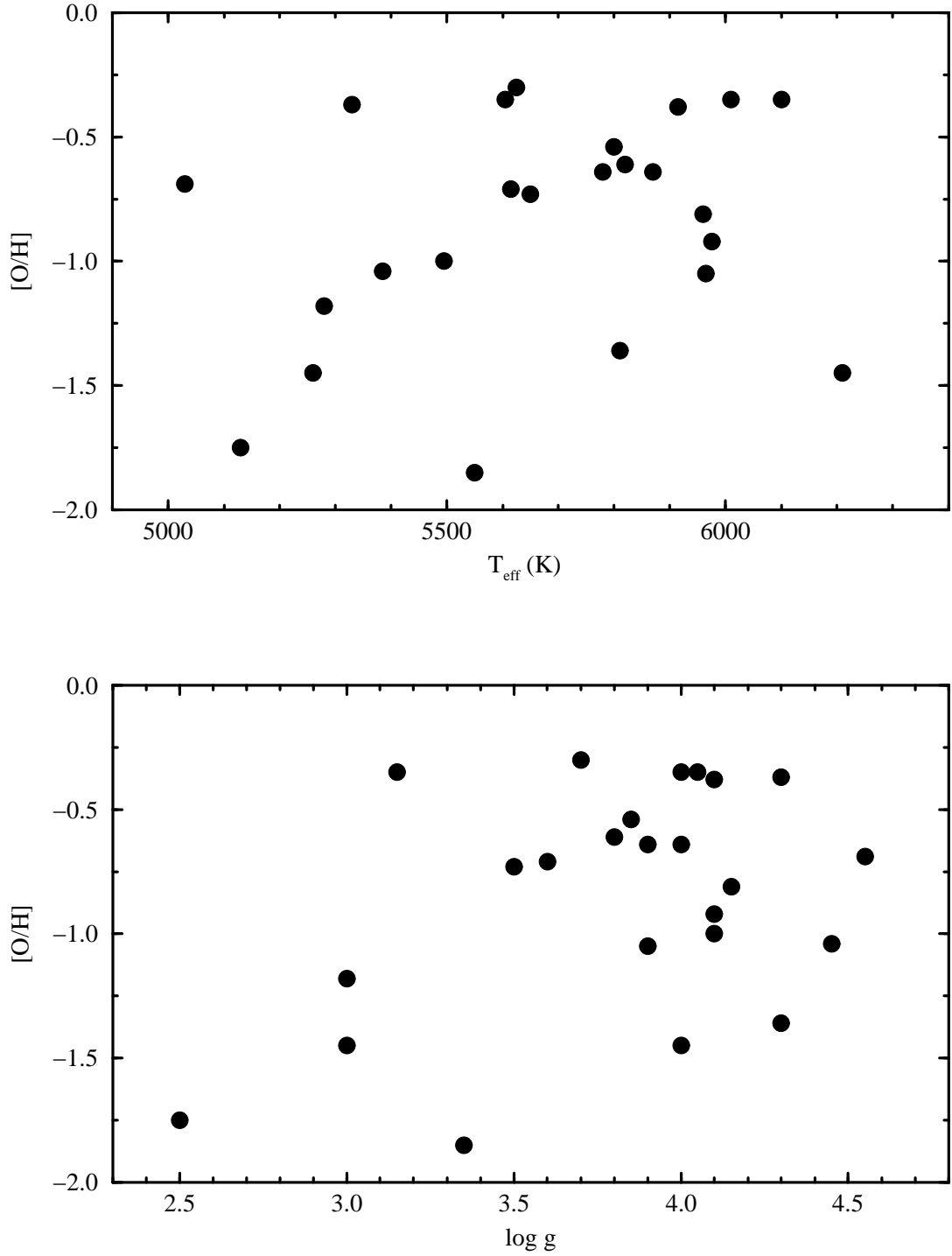}
\caption[Fig6.ps]{Oxygen abundances against adopted effective temperatures
(upper panel) and surface gravities (lower panel), showing no dependence of the
abundances derived from the OH lines on these parameters. \label{fig6}}
\end{figure}

\begin{figure}[ht]
\plotone{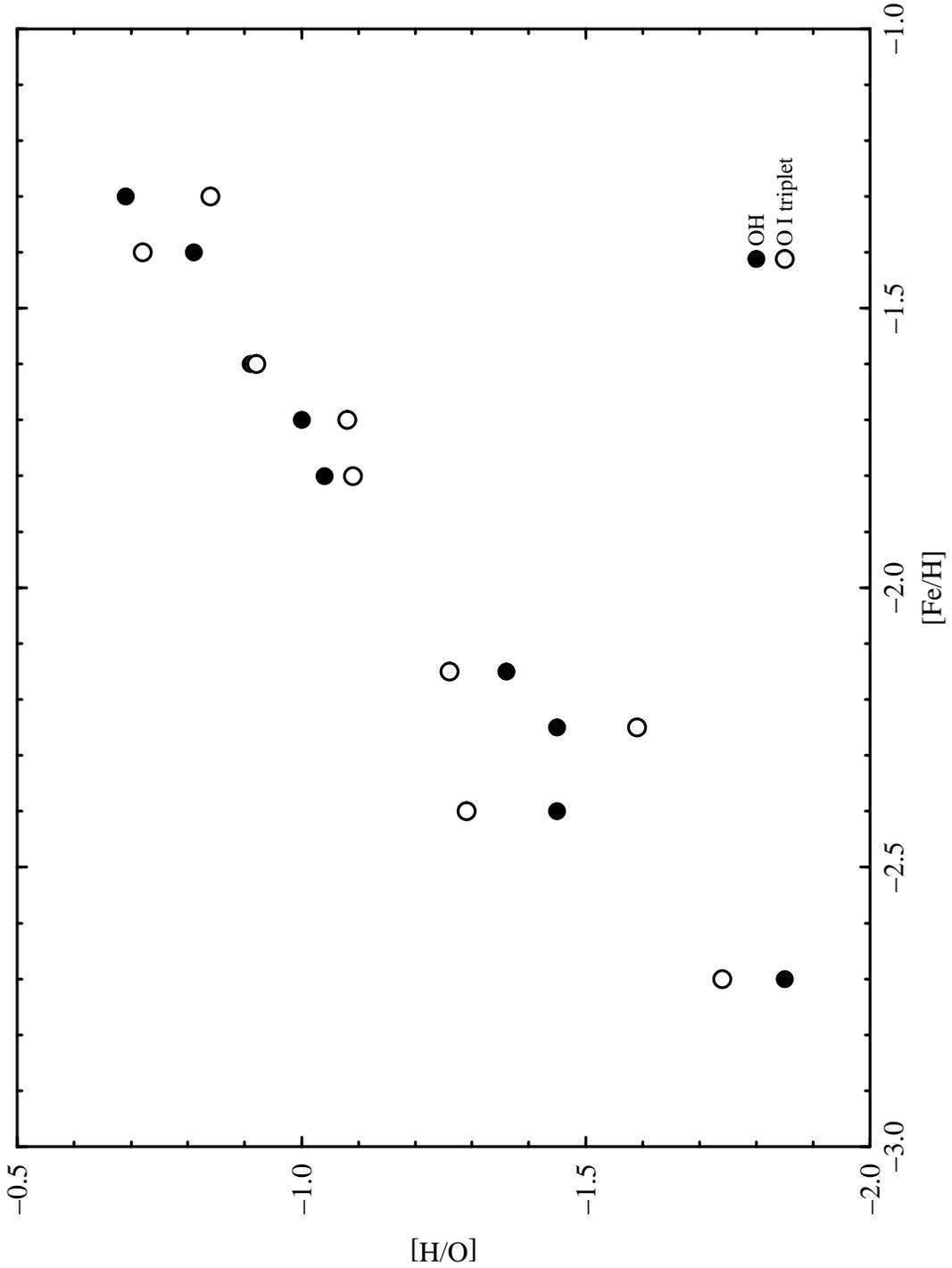}
\caption[Fig7.ps]{Comparison between oxygen abundances derived from OH lines
and the IR \ion{O}{1} triplet for nine stars  common to our sample and that of 
Tomkin et al. (1992). The stellar parameters listed in Table 1, and the
equivalent widths of the \ion{O}{1} triplet provided by Tomkin et al. were used
to carry out an LTE abundance analysis. There are no systematic trends between
both set of abundances, with a mean difference of $0.00\pm 0.11$
dex.\label{fig7}}
\end{figure}

\begin{figure}[hb]
\plotone{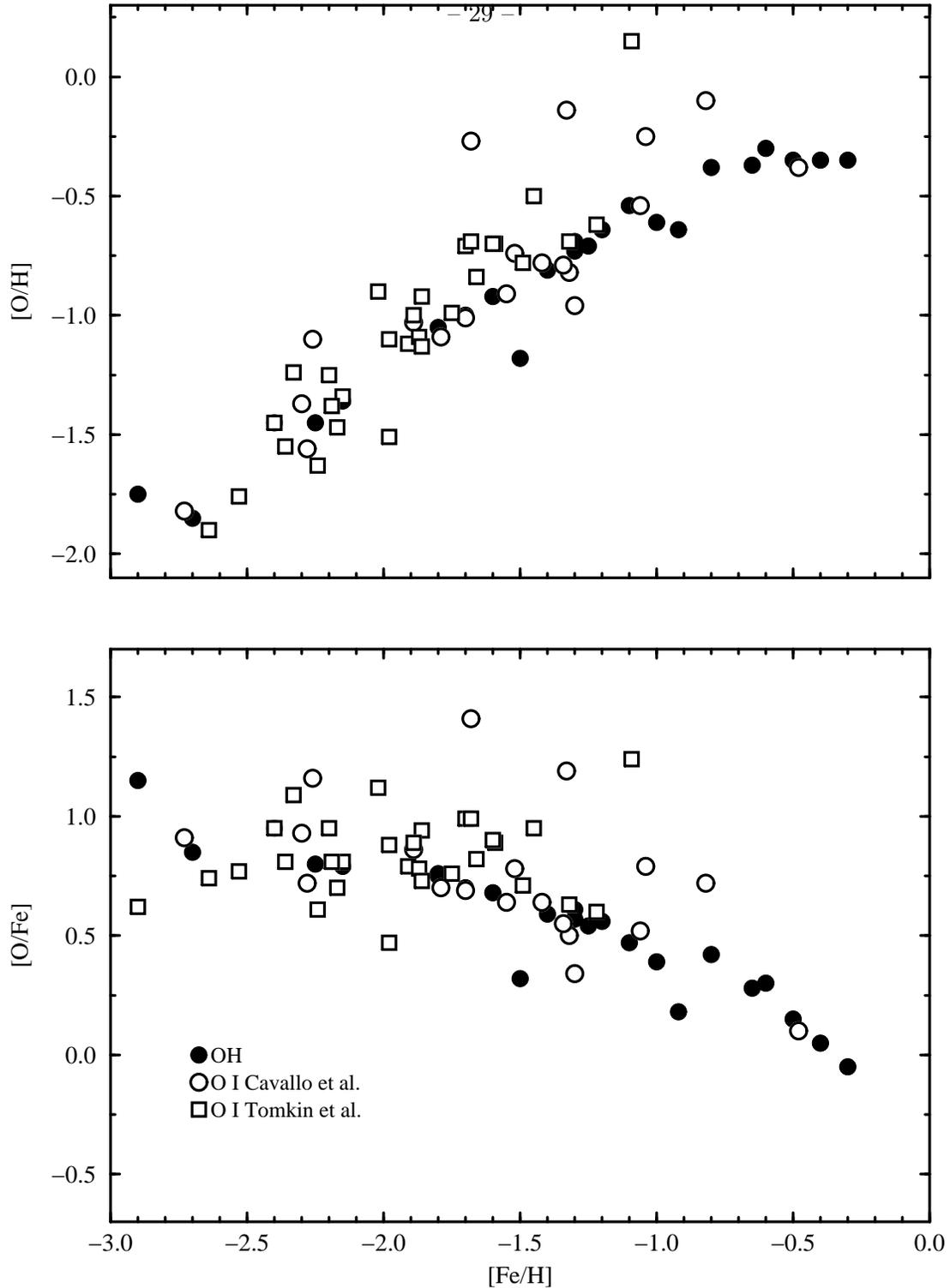}
\caption[Fig8.ps]{Comparison between the trends against metallicity delineated
by oxygen abundances ([O/H] in the upper panel, and [O/Fe]  in the lower panel)
derived from OH lines and the IR \ion{O}{1} triplet. LTE abundances associated
with the triplet were taken directly from  Tomkin et al. (1992) and Cavallo et
al. (1997). The oxygen abundances derived from OH lines tend to conform a lower
envelope to the observed trends.\label{fig8}}
\end{figure}

\clearpage

\begin{deluxetable}{lccccc}
\footnotesize
\tablecaption{Stellar parameters of the program stars and final oxygen
abundances.
\label{tbl-1}}
\tablewidth{0pt}
\tablehead{
\colhead{Star} & \colhead{$T_{\rm eff}$} & \colhead{$\log g$} & \colhead{[Fe/H]}
& \colhead {[O/H]}& \colhead{[O/Fe]} 
}
\startdata 
HD 6582    & 5330$\pm$100 & 4.30$\pm$0.15 & $-$0.65$\pm$0.10 & $-$0.37$\pm$0.22 &\phs0.28 \nl
HD 19445   & 5810$\pm$150 & 4.30$\pm$0.10 & $-$2.15$\pm$0.10 & $-$1.36$\pm$0.32 &\phs0.79 \nl
HD 22879   & 5915$\pm$100 & 4.10$\pm$0.15 & $-$0.80$\pm$0.20 & $-$0.38$\pm$0.23 &\phs0.42 \nl
HD 38510   & 5780$\pm$100 & 3.90$\pm$0.10 & $-$0.92$\pm$0.20 & $-$0.72$\pm$0.22 &\phs0.20 \nl
HD 64090   & 5385$\pm$100 & 4.45$\pm$0.10 & $-$1.80$\pm$0.10 & $-$1.04$\pm$0.22 &\phs0.76 \nl
HD 76932   & 5800$\pm$100 & 3.85$\pm$0.10 & $-$1.10$\pm$0.10 & $-$0.62$\pm$0.23 &\phs0.48 \nl
HD 84937   & 6210$\pm$120 & 4.00$\pm$0.10 & $-$2.25$\pm$0.15 & $-$1.45$\pm$0.25 &\phs0.80 \nl
HD 94028   & 5975$\pm$120 & 4.10$\pm$0.10 & $-$1.60$\pm$0.10 & $-$0.91$\pm$0.25 &\phs0.69 \nl
HD 103095  & 5030$\pm$100 & 4.55$\pm$0.35 & $-$1.30$\pm$0.10 & $-$0.68$\pm$0.26 &\phs0.62 \nl
HD 134169  & 5820$\pm$100 & 3.80$\pm$0.15 & $-$1.00$\pm$0.20 & $-$0.61$\pm$0.22 &\phs0.39 \nl
HD 140283  & 5550$\pm$100 & 3.35$\pm$0.10 & $-$2.70$\pm$0.10 & $-$1.85$\pm$0.22 &\phs0.85 \nl
HD 157214  & 5625$\pm$100 & 3.70$\pm$0.15 & $-$0.60$\pm$0.20 & $-$0.29$\pm$0.22 &\phs0.31 \nl
HD 165908  & 6010$\pm$120 & 4.05$\pm$0.15 & $-$0.40$\pm$0.20 & $-$0.35$\pm$0.25 &\phs0.05 \nl
HD 166913  & 5965$\pm$120 & 3.90$\pm$0.20 & $-$1.80$\pm$0.10 & $-$1.05$\pm$0.27 &\phs0.75 \nl
HD 170153  & 6100$\pm$120 & 4.00$\pm$0.20 & $-$0.30$\pm$0.20 & $-$0.35$\pm$0.26 & $-$0.05 \nl
HD 188510  & 5495$\pm$100 & 4.10$\pm$0.10 & $-$1.70$\pm$0.10 & $-$1.01$\pm$0.23 &\phs0.69 \nl
HD 189558  & 5650$\pm$100 & 3.50$\pm$0.10 & $-$1.30$\pm$0.20 & $-$0.73$\pm$0.22 &\phs0.57 \nl
HD 194598  & 5960$\pm$120 & 4.15$\pm$0.10 & $-$1.40$\pm$0.15 & $-$0.81$\pm$0.27 &\phs0.59 \nl
HD 201889  & 5615$\pm$100 & 3.60$\pm$0.15 & $-$1.25$\pm$0.15 & $-$0.71$\pm$0.25 &\phs0.54 \nl
HD 201891  & 5870$\pm$100 & 4.00$\pm$0.10 & $-$1.20$\pm$0.20 & $-$0.63$\pm$0.23 &\phs0.57 \nl
HD 211998  & 5280$\pm$100 & 3.00$\pm$0.20 & $-$1.50$\pm$0.15 & $-$1.18$\pm$0.23 &\phs0.32 \nl
HD 225239  & 5605$\pm$100 & 3.15$\pm$0.20 & $-$0.50$\pm$0.20 & $-$0.36$\pm$0.22 &\phs0.14 \nl
BD $+$23\arcdeg 3130 & 5130$\pm$150 & 2.50$\pm$0.30 & $-$2.90$\pm$0.20 & $-$1.73$\pm$0.33 &\phs1.17 \nl
BD $+$37\arcdeg 1458 & 5260$\pm$100 & 3.00$\pm$0.20 & $-$2.40$\pm$0.10 & $-$1.43$\pm$0.27 &\phs0.97 \nl
                      
\enddata
\end{deluxetable}

\clearpage

\begin{deluxetable}{lcccccc}
\footnotesize
\tablecaption{Selected OH (0,0) lines\tablenotemark{a} ~used for the abundance 
analysis. 
\label{tbl-2}}
\tablewidth{0pt}
\tablehead{
\colhead{Transition} & \colhead{$\lambda_{\rm lab}$ (\AA)} & 
\colhead{$\lambda_\odot$ (\AA)} & \colhead{$\log gf_{\rm th}$} &
\colhead{$\log gf_{\rm ad}$} & \colhead{$\Delta \log gf$} & 
\colhead{$\chi$ (eV)}
}
\startdata             
$R_{22}$(18.5) & 3084.895 & 3084.887 &$-$1.943&$-$1.906 &$-$0.037 &0.067\nl
$Q_{11}$(5.5)  & 3085.199 & 3085.194 &$-$2.013&$-$2.060 &$-$0.007 &0.843\nl
$P_{11}$(9.5)  & 3123.948 & 3123.945 &$-$1.982&$-$2.086 &$+$0.104 &0.204\nl
$Q_{22}$(15.5) & 3127.687 & 3127.672 &$-$1.581&$-$1.590 &$+$0.009 &0.612\nl
$R_{22}$(5.5)  & 3128.060 & 3128.060 &$-$2.425&$-$2.358 &$-$0.067 &0.102\nl
$Q_{12}$(8.5)  & 3128.107 & 3128.101 &$-$2.998&$-$3.221 &$+$0.223 &0.209\nl
$P_{22}$(8.5)  & 3128.286 & 3128.278 &$-$2.026&$-$2.070 &$+$0.044 &0.209\nl
$Q_{22}$(17.5) & 3139.169 & 3139.161 &$-$1.559&$-$1.762 &$+$0.203 &0.760\nl
$Q_{22}$(21.5) & 3167.169 & 3167.167 &$-$1.823&$-$1.694 &$-$0.129 &1.108\nl
$P_{22}$(16.5) & 3186.084 & 3186.088 &$-$1.958&$-$2.097 &$+$0.139 &0.685\nl
$P_{22}$(18.5) & 3203.975 & 3203.970 &$-$1.815&$-$1.920 &$+$0.105 &0.843\nl
$P_{22}$(23.5) & 3255.492 & 3255.485 &$-$1.844&$-$1.835 &$-$0.009 &1.300\nl 
\enddata
\tablenotetext{a}{The line $\lambda$ 3128.06 \AA\ belongs to the (1,1) band.}
\end{deluxetable}

\clearpage

\begin{deluxetable}{lrrrrrrrrrrrrc}
\scriptsize
\tablecaption{Oxygen abundances derived from the OH lines listed in
Table 2. \label{tbl-3}}
\tablewidth{0pt}
\tablehead{
\colhead{Star} & \colhead{3084}   & \colhead{3085}   & 
\colhead{3123} &
\colhead{3127}  & \colhead{3128.1 \tablenotemark{a}} & \colhead{3128.3} &
\colhead{3139} & \colhead{3167} &
\colhead{3186 \tablenotemark{b}} & \colhead{3203} & \colhead{3255} & \colhead{[O/H]}
}
\startdata
HD 6582    & --  & --  &$-$0.35 &$-$0.45 &$-$0.35 &$-$0.45 & --  & --  &$-$0.25 & --  & -- & $-$0.37$\pm$0.08\nl
HD 19445   & $-$1.35 & $-$1.25 & $-$1.35 & $-$1.45 & $-$1.35 & $-$1.45 & $-$1.25 & $-$1.35 & $-$1.25 & $-$1.35 & $-$1.55 &$-$1.36$\pm$0.09\nl
HD 22879   & --  & --  &$-$0.40 & --  &$-$0.45 &$-$0.20 & $-$0.45 & -- & -- & --  & -- &$-$0.38$\pm$0.10 \nl
HD 38510   & --  & --  &$-$0.60 & $-$0.80 & $-$0.70 & $-$0.70 & $-$0.80 & -- & -- & -- & -- & $-$0.72$\pm$0.08\nl
HD 64090   & --  & --  & $-$0.90 & $-$1.00 & $-$1.10 & $-$1.10 & $-$1.10 & -- & -- & -- & -- & $-$1.04$\pm$0.08\nl
HD 76932   & $-$0.70 & $-$0.50 & $-$0.50 & $-$0.50 & $-$0.65 & $-$0.50 & $-$0.70 & $-$0.70 & $-$0.60 & $-$0.70 &$-$0.80 &$-$0.62$\pm$0.10 \nl
HD 84937   & --  & --  & $-$1.40 & $-$1.55 & -- & $-$1.50 & $-$1.45 & $-$1.35 & -- & -- & -- & $-$1.45$\pm$0.07\nl
HD 94028   & --  & --  & $-$0.80 & --  & $-$0.90 & $-$0.90 & $-$0.90 & $-$0.90 & $-$1.00 & -- & $-$1.00 & $-$0.91$\pm$0.06\nl
HD 103095  & --  & --  & $-$0.60 & $-$0.60 & $-$0.60 & $-$0.60 & $-$0.80 & $-$0.90 & & & & $-$0.68$\pm$0.12\nl
HD 134169  & $-$0.50 & $-$0.60 & $-$0.50 & $-$0.60 & $-$0.60 & $-$0.60 & $-$0.55 & $-$0.70 & $-$0.60 & $-$0.60 &$-$0.80 &$-$0.61$\pm$0.08\nl
HD 140283  & $-$1.80 & $-$1.80 & $-$1.80 & $-$1.90 & $-$1.80 & $-$1.80 & $-$1.80 & $-$1.95 & $-$1.80 & $-$1.95 & $-$1.95 &$-$1.85$\pm$0.07 \nl
HD 157214  & $-$0.40 &$-$0.30 &$-$0.20 &$-$0.30 &$-$0.30 &$-$0.30 &$-$0.30 & $-$0.20 &$-$0.20 &$-$0.30 &$-$0.40&$-$0.29$\pm$0.07\nl
HD 165908  & $-$0.40 &$-$0.30 &$-$0.30 &$-$0.40 &$-$0.30 &$-$0.40 &$-$0.25 &$-$0.40 &$-$0.30 &$-$0.40 &$-$0.40&$-$0.35$\pm$0.06 \nl
HD 166913  & --  & --  & $-$1.15 & -- & -- & -- & $-$0.95 & -- & -- & -- & -- & $-$1.05$\pm$0.10 \nl
HD 170153  & $-$0.40 &$-$0.30 &$-$0.30 &$-$0.40 &$-$0.20 &$-$0.50 &$-$0.30 &$-$0.40 &$-$0.30 &$-$0.40 &$-$0.35 & $-$0.35$\pm$0.08\nl
HD 188510  & --  & --  & --  &$-$1.20 & $-$1.00 & $-$1.00 & $-$0.95 & $-$0.90 & $-$0.90 & $-$1.00 & $-$1.10 &$-$1.01$\pm$0.10 \nl
HD 189558  & --  & --  & $-$0.70 & $-$0.60 & $-$0.60 & $-$0.80 & $-$0.70 & $-$0.80 & $-$0.70 & $-$0.80 & $-$0.85 & $-$0.73$\pm$0.09\nl
HD 194598  & --  & --  & $-$0.95 & $-$0.80 & $-$0.70 & $-$1.00 & $-$0.80 & $-$0.70 & $-$0.70 & -- & $-$0.80 &$-$0.81$\pm$0.11\nl
HD 201889  & --  & --  & $-$0.85 & $-$0.65 & $-$0.75 & $-$0.75 & $-$0.55 & $-$0.55 & $-$0.55 & $-$0.80 &$-$0.95 &$-$0.71$\pm$0.14\nl
HD 201891  & --  & --  & $-$0.50 & $-$0.70 & $-$0.65 & $-$0.60 & $-$0.60 & $-$0.60 & $-$0.50 & $-$0.75 & $-$0.80 & $-$0.63$\pm$0.10\nl
HD 211998  & --  & --  &  $-$1.30 & $-$1.20 & $-$1.10 & $-$1.20 & $-$1.10 & -- & -- & -- & -- & $-$1.18$\pm$0.08\nl
HD 225239  & $-$0.40 &$-$0.30 &$-$0.30 &$-$0.40 &$-$0.35 &$-$0.30 &$-$0.30 & $-$0.35 &$-$0.30 &$-$0.45 &$-$0.45&$-$0.36$\pm$0.06 \nl
BD $+$23\arcdeg 3130 & --  & --  & $-$1.70 & $-$1.70 & $-$1.70 & $-$1.80 & $-$1.60 & $-$1.70 & $-$1.70 & $-$1.90 & -- &$-$1.73$\pm$0.08\nl
BD $+$37\arcdeg 1458 & --  & --  & $-$1.20 & $-$1.30 & $-$1.50 & $-$1.20 & $-$1.50 & $-$1.50 & $-$1.40 & $-$1.55 & $-$1.70 & $-$1.43$\pm$0.16\nl

\enddata

\tablenotetext{a}{Blended with OH (1,1) 3128.06 \AA.}
\tablenotetext{b}{Blended with OH (1,1) 3185.98 \AA.}

\end{deluxetable}

\clearpage

\begin{deluxetable}{lcccccccc}
\scriptsize
\tablecaption{Comparison between oxygen abundances derived from the OH and
[O I] lines.\label{tbl-4}}
\tablewidth{0pt}
\tablehead{
\colhead{Star} & \colhead{Ref.} & 
\colhead{$T_{\rm eff}$/$\log g$/[Fe/H]} & 
\colhead{[O/H]$_{\rm [O I]}$} &
\colhead{[O/H]$_{\rm [O I]}$} & 
\colhead{[O/H]$_{\rm OH}$}   &
\colhead{$\log g$} & 
\colhead{[O/H]$_{\rm OH}$} &
\colhead{[O/H]$_{\rm [O I]}$} \nl
& & literature & literature & Table 1 & Table 1 & {\it Hipparcos} & 
{\it Hipparcos} & {\it Hipparcos} 
}
\startdata          
HD 22879  & K\&B & 5859/4.29/$-$0.84 & $-$0.59  & $-$0.61 & $-$0.38 & 4.38 & $-$0.48 & $-$0.50 \nl
HD 76932  &  C   & 5840/3.90/$-$1.04 & $-$1.02  & $-$1.00 & $-$0.62 & 4.12 & $-$0.72 & $-$0.90 \nl
HD 103095 & B\&C & 5050/4.70/$-$1.22 & $-$0.89  & $-$0.93 & $-$0.68 & 4.68 & $-$0.73 & $-$0.88 \nl
HD 134169 & S\&S & 5780/3.40/$-$1.00 & $-$0.97  & $-$0.78 & $-$0.61 & 4.00 & $-$0.68 & $-$0.70 \nl
\enddata

\tablerefs{K\&B, King \& Boesgaard 1995; C, Cavallo et al. (1997); B\&C,
Balachandran \& Carney (1996); S\&S, Spite \& Spite (1991). For HD 76932 we
list the stellar parameters and oxygen abundance given by Cavallo et al. 
(1997) who reanalyzed the measurement of Magain (1987), while for HD 103095 we
show the parameters and abundance listed by Balachandran \& Carney (1996) who
used the equivalent width provided by Spite \& Spite (1991).}

\end{deluxetable}

\end{document}